\documentclass[11pt,a4paper]{article}

\usepackage{amsmath}
\usepackage{amssymb}

\newcommand{\eq}[1]{\begin{align*} #1 \end{align*}}
\newcommand{\bs}[1]{\mathbf{#1}}

\title{Errata list for ``Error Control Coding'' \\ by Lin and Costello}
\author{Erik Agrell\\
Department of Electrical Engineering,\\
Chalmers University of Technology,\\
G\"oteborg, Sweden\\
agrell@chalmers.se
}
\date{v.~2.10, Mar.~13, 2021}

\begin{document}

\maketitle

\section*{Introduction}

This document lists some errors found in the second edition of \emph{Error Control Coding} by Shu Lin and Daniel J.~Costello, Jr.~\cite{lin2004}. The list is made in good faith, in the hope that it will be both correct and helpful, but it is not endorsed by the book's authors or publisher.

Typographical text errors, where the meaning is obvious, are omitted. On the other hand, some reading hints, which do not exactly indicate errors, are included.

The author of this list would like to thank Alex Alvarado, Frank Bologna, Arash Ghasemmehdi, Naga VishnuKanth Irukulapati, Clay McKell, Leela Srikar Muppirisetty, Mohsen Nosratinia, Alireza Sheikh, Arash Tahmasebi Toyserkani, and Mehrnaz Tavan for contributions. He would appreciate reports of other errors in the book or in this list.

\section*{Errata}

\begin{itemize}

\item p.~4, lines 5--6 from below: With the usual definition of redundancy, it is not correct that ``more redundancy is added by increasing the memory order $m$ while holding $k$ and $n$, and hence the code rate $R$, fixed''. A correct statement would be that \emph{lower error probabilities} are achieved by increasing the memory order $m$ while holding $k$ and $n$ fixed, as mentioned on p.~453.

\item p.~9, (1.6): $y+\sqrt{E_s}$ should be $(y+\sqrt{E_s})^2$ and $y-\sqrt{E_s}$ should be $(y-\sqrt{E_s})^2$.

\item p.~10, line 6: $\frac{1}{2}T$ should be $\frac{1}{2T}$.

\item p.~12, lines 8--12 after (1.12): $p(r_i | v_i)$ should be $P(r_i | v_i)$ and ``probability $P(\bs{r} | \bs{v})$'' should be ``probability density function $p(\bs{r} | \bs{v})$'' twice.

\item p.~12, last line: ``actually on the average'' should be ``actually the average''.

\item p.~18, (1.21): The right-hand side is meant to be interpreted as
\eq{
10 \log_{10} \left( \frac{1}{2} R d_\mathit{min} \right).
}
This expression, however, is still not accurate for odd values of $d_\mathit{min}$ (with $d_\mathit{min}=1$ being the most obvious example). The correct expression is
\eq{
10 \log_{10} \left(R \left\lceil \frac{d_\mathit{min}}{2} \right\rceil \right),
}
where $\lceil x \rceil$ denotes the smallest integer not less than $x$.

\item pp.~19 and 21: ``BER'' in the legends of Figures 1.11 and 1.12 should be ``BPSK coded''.

\item p.~20, line 3: 3.91 should be 3.89. The entries of Table~1.2 differ by up to 0.014 dB from the exact values.

\item p.~21, line 6 from below: $R/W$ should be $(R/T)/W$, because the information transmission rate in bps is denoted by $R/T$ in this book (see \cite[p.~10]{lin2004}), not $R$ as in \cite[p.~2]{proakis} (which is ref.~7 in \cite[p.~21]{lin2004}).

\item p.~38, line 17: The word ``uniquely'' is missing before ``solve the preceding equations''.

\item p.~40, lines 10--24: The theorem that $f(X)=q(X)g(X)+r(X)$ has a unique solution $(q(X),r(X))$ for given $(f(X),g(X))$, where the degree of $r(X)$ is less than the degree of $g(X)$, is usually not called ``Euclid's division algorithm''. The purpose of Euclid's division algorithm, often simply called the Euclidean algorithm, is to find the greatest common divisor of two integers or polynomials, see Section 7.5.

\item p.~45, (2.19c): $(a_{i0}+a_{i1}+\cdots+a_{i,m-1}\alpha^{m-1})$ should be $(a_{i0}+a_{i1}\alpha+\cdots+a_{i,m-1}\alpha^{m-1})$.

\item p.~53, line 3 from below: $GF(2^7)$ should be $GF(2^4)$.

\item p.~64, Problem 2.23: ``vector space $\mathit{GF}(2)$'' should be ``vector space over $\mathit{GF}(2)$''.

\item p.~67, (3.3): $\bs{u}\cdot G$ should be $\bs{u}\cdot\bs{G}$.

\item p.~81, line 2: ``closer to \ldots than'' should be ``at least as close to \ldots as''.

\item p.~89, Figure~3.8: $s_{n-1}$ should be $s_{n-k-1}$.

\item p.~89, Example 3.9: $\Lambda$ should be $\wedge$ throughout the example.

\item p.~97, ref.~3: J.~J.~A. Sloane should be N.~J.~A. Sloane.

\item p.~130, line 6: ``columns and rows'' should be ``rows and columns''.

\item p.~130, line 9: $[(d_1d_2-1)/2]$ should be $\lfloor(d_1d_2-1)/2\rfloor$.

\item pp.~131, 133: ``$d_1+d_2-1$'' on p.~131 and in Problem~4.22 should be ``at least $d_1+d_2-1$''. The bound holds with equality if and only if both codes contain a minimum-weight codeword with only one nonzero information bit.

\item p.~134, ref.~7: T.~Klove should be T.~Kl\o ve and V.~I. Korzhik.

\item p.~142, line 11: ``followed by'' should be ``preceded by''.

\item p.~143, last line of Table~5.2: $1+X^2+X^5$ should be $1+X^2+X^3$.

\item p.~147, caption of Figure~5.1: $g_1 X^2$ should be $g_1 X$.

\item p.~151, line 1 after (5.21): ``Dividing both sides of (5.21) by $\bs{g}(X)$'' should be removed.

\item p.~190, Problem~5.12: $e^{(i)}(X)$ should be $\bs{e}^{(i)}(X)$.

\item p.~195: It is stated after (6.3) that if $t$ is small, then $n-k=mt$. In this context, ``small'' should be interpreted in relation to $m$. For example, if $t=3$, then $n-k=mt$ for $m\ge 5$ but not for $m=4$.

\item pp.~196--197, Table~6.1: The top rows on p.~196 are for $n=63$, $255$, and $511$, resp. The top rows on p.~197 are all for $n=1023$. For $n=255$ and $k=139$, $131$, $123$, $115$, $107$, $99$, the $t$ values should be $15$, $18$, $19$, $21$, $22$, $23$, resp. For $(n,k)=(511,10)$, $t=127$, and for $(n,k)=(1023,16)$, $t=247$.

\item p.~211,  (6.28): $\bs{b}_5(X) = X^2$ should be $\bs{b}_5(X) = 1+X$. This error does not propagates to the rest of Example 6.5 and does not change the result, since now $S_5 = 1+\alpha^5 = \alpha^{10}$, which is the same value as in the unnumbered equation after (6.29).

\item p.~233, ref.~25: Sidel\'nikov should be Sidel'nikov.

\item p.~238, (7.3): $(q^{2t}-q^i)$ should be $(q^{2t}-q^j)$.

\item pp.~344--355, 361, and 373: The symbol $\phi$ (phi) should, wherever it is \emph{not} followed by a parenthesis, be replaced with $\varnothing$ (empty set). The function $\phi(\cdot)$, however, denotes an interval (pp.~344--345 and 372--373) and should be kept unchanged.

\item p.~439, last paragraph: Section 4.2 should be Section 4.3 twice.

\item p.~461, (11.26a): $[\bs{v}^0(D),\bs{v}^1(D)]$ should be $[\bs{v}^{(0)}(D),\bs{v}^{(1)}(D)]$.

\item p.~466, 2--3 lines before (11.47): $h_i^{(j)}$ should be $\bs{h}_i^{(j)}$ twice.

\item p.~483: $\stackrel{>}{=}$ should be $\ge$ in three places.

\item p.~487, (11.95): $\boldsymbol{\sigma}_k$ should be $\boldsymbol{\sigma}_l$.

\item p.~487, (11.96): $\boldsymbol{\sigma}$ should be $\boldsymbol{\sigma}_l$.

\item p.~493, line 1 after (11.107): ``Problem~11.8'' should be ``Problem 11.17''.

\item p.~499, last line before (11.129): ``any $w$ and $d$'' should be ``any $d$''.

\item p.~500, (11.134): The last sum should be over $w$ only, not $z$.

\item p.~511: If Problem~11.3b was intended to agree with Problem~11.1c, then $\bs{u}(D)$ should be $1+D+D^2+D^4$.

\item p.~517, line 3 after (12.3): $\log(\bs{r}|\bs{v})$ should be $\log P(\bs{r}|\bs{v})$.

\item p.~519: $c_2 = 17.3$ should be approximately $16.7$. Although both values generate Fig.~12.4(b) after rounding to the nearest integer, $c_2=17.3$ is hardly ``chosen so that all metrics can be approximated by integers''.

\item p.~522, last line before (12.9): (1.11) should be (1.12).

\item p.~525, second line of (12.15): $(1+X^2L^2)$ should be $(1+X^2L)^2$.

\item p.~526, (12.16): $\binom{7}{3}$ should be $\binom{7}{e}$.

\item p.~528, Figure~12.9: The figure illustrates three paths: $\bs{v}$ which is the lowest path, $\bs{v}''$ which is the highest path, and $\bs{v}'$ which goes first high and then low.

\item p.~532, (12.38): The expression is not correct for odd $d_\mathit{free}$, because the inequality on the second line of (12.21) is not asymptotically tight as $p \rightarrow 0$. The correct expression is
\eq{
\gamma \triangleq 10\log_{10}\left(R\left\lceil\frac{d_\mathit{free}}{2}\right\rceil\right)\text{ dB}
}

\item p.~554, line 14: 1/3 should be 3.

\item pp.~559--560: $[\bs{r}|\bs{v}]_t$ should be $[\bs{r}|\bs{v}]_{t+1}$ and $[\bs{r}|\bs{v}]_{t-1}$ should be $[\bs{r}|\bs{v}]_t$ in (12.86), (12.89), (12.94), and two lines after (12.94), to agree with the definition (12.5).

\item p.~559, (12.89): $u_t$ means two different things in the last line of this equation; a random variable and a value that this random variable may assume. If we denote with $U_t$ a random variable that represents the unknown information bit at time $t$, then
\eq{
2 \ln P(U_t=b) - C_u = b \ln \frac{P(U_t = +1)}{P(U_t = -1)}
}
for any $b \in \{-1,1\}$, which is the last term of (12.89).

\item pp.~559--560: This is not an error but it can be easily misunderstood. $L(u)$ in (12.91) is a constant, \emph{not} a function of $u$. (I.e., $L(1)$ and $L(-1)$ do not make sense.) Similarly, $L(u_t)$ in (12.94) is possibly a function of $t$ but not of $u_t$.

\item p.~561, (12.99): $c$ should be $c/2$ in three places.

\item p.~564: The a priori L-values $L_a$, which are introduced without a definition, should be defined in analogy with (12.91) as
\eq{
L_a(u_l) = \ln\left[\frac{p(u_l=+1)}{p(u_l=-1)}\right]
}
For any given $l$, $L_a(u_l)$ is a constant, not a function of $u_l$.

The a posteriori L-values $L(u_l)$ defined in (12.106) depend on $\bs{r}$ and $l$ but not on the value of $u_l$.

\item p.~562: Equation (12.104) is not correct. If we rewrite (12.103) as
\eq{
\bs{L}_t(S_i) = [L_0^t(S_i),L_1^t(S_i),\cdots,L_{t-1}^t(S_i)]
}
to make the time dependence explicit, then (12.104) should read
\eq{
L_l^t(S_i) = \begin{cases}
\min[\Delta_{t-1}(S_i),\,L_l^{t-1}(S_i')] & \textrm{if }u_l \ne u_l'\\
L_l^{t-1}(S_i') & \textrm{if }u_l = u_l'\\
\end{cases}
,\, l=0,1,\cdots,t-1
}
where $S_i'$ is the state through which the ML path to $S_i$ passes at time $t-1$. By definition, $L_{t-1}^{t-1}(S)=\infty$ for all $S$. With this notation, the top line of Figure~12.18, which represents $\bs{L}_t(S_1)$, should read
\eq{
L_{t-6}^{t-1}(S_0)~~%
\begin{array}{c}
\min\{\Delta_{t-1}(S_1),\\
L_{t-5}^{t-1}(S_0)\}
\end{array}~~%
L_{t-4}^{t-1}(S_0)~~%
\begin{array}{c}
\min\{\Delta_{t-1}(S_1),\\
L_{t-3}^{t-1}(S_0)\}
\end{array}~~%
L_{t-2}^{t-1}(S_0)~~%
\infty
}

\item $P(\bs{r})$ on pp.~564--565 should be read as $p(\bs{r})$, because it is a probability density function in the soft-decision case.

\item p.~567, (12.123): $u_l$ means two different things in this equation; a random variable and a value that this random variable may assume. What (12.123) tries to say is, with a different notation, $P(U_l=b) = A_l e^{b L_a^{(l)}/2}$, where $b \in \{-1,1\}$, $U_l$ is a random variable representing the unknown information bit at time $l$,
\eq{
A_l = \sqrt{P(U_l=1) P(U_l=-1)},
}
and
\eq{
e^{L_a^{(l)}/2} = \sqrt{\frac{P(U_l=1)}{P(U_l=-1)}}
.}
Both $A_l$ and $e^{L_a^{(l)}/2}$ depend on the statistics of $U_l$, but none of them depend on a certain outcome $b$.

\item p.~576, (12.142): The expressions should be
\eq{
  \beta_1^\ast (S_0) &= \max(1.00,3.30) = 3.30 \\
  \beta_1^\ast (S_1) &= \max(2.00,2.30) = 2.30
}

\item p.~576, (12.143): The last line in the expression for $L(u_0)$ should be
\eq{
  = (2.75)-(2.85) = -0.10
}
and the last two lines in the expression for $L(u_1)$ should be
\eq{
  &= \max[(2.45),(2.85)]-\max[(0.55),(2.75)] \\
  &= (2.85)-(2.75) = +0.10
}

\item pp.~585--586: The concept of ``tail-biting convolutional code'' is used incorrectly in parts of Section 12.7. One way (not the only way) to fix this is to replace ``tail-biting'' with ``unterminated'' twice on p.~585 and once on the first line of p.~586. Also replace ``modify'' with ``introduce'' on p.~586, line 14. No changes are needed in the rest of the section, where ``tail-biting'' is used in its usual sense, which is the sense defined in the last paragraph of Section 11.1, p.~486.

\item p.~598, Problem~12.1: ``code listed in Table~12.1(d)'' should probably be ``encoder in Example 11.2'', because a given code can have several different trellises depending on the encoder implementation.

\item p.~599, Problem~12.10: The value $p=0.1$ in Problem~12.10a should be reduced, because the bounds (12.25) and (12.29) diverge for $p \ge 0.055$. It seems plausible that $p=0.01$ and $0.001$ were intended in the two subproblems, because these values were used in the first edition of the book \cite[Problem~11.8]{lin1983}.

\item p.~689, ref.~23: The first two authors, Vinck and de Paepe, should be swapped.

\item p.~739, lines 5--7: The sentence ``Concatenated coding\ldots'' is incomplete. Perhaps the last comma should be replaced with ``and the''.

\item p.~754, lines 8--9: The sentence ``In the following\ldots'' is incomplete. Probably ``is described'' or something similar should be appended.

\item p.~763, Problem~15.2: 15.12 should be (15.12).

\item p.~766, line 3 from below: ``exceed'' is incorrect in this context, because the discussed $E_b/N_0$ value is \emph{less} (and hence better) than that of previously known codes.

\item p.~768, Figure~16.1(b): The arrows for $\bs{v}^{(1)}$ and $\bs{v}^{(2)}$ should both be moved 1~cm down, to represent the outputs of the adders.

\item p.~822, (16.103a): The denominator $P_{L_a}(\xi \,| u_l=-1) + P_{L_a}(\xi \,| u_l=+1)$ should be understood as $P_{L_a}(\xi \,| {-1}) + P_{L_a}(\xi \,| {+1})$, because it is not a function of the summation variable $u_l$.

\item p.~830, line 3 after (16.108a): (16.107a) should be (16.108a).

\item p.~832, (16.114b): $-\frac{1}{2}(L_{u0}+L_{p0})$ should be $\frac{1}{2}(L_{u0}+L_{p0})$.

\item pp.~875--879: The backslash symbol is, by mathematical convention, defined as an operation on two sets (the first set minus the second). Therefore, the notation $X \setminus y$ should be read as $X \setminus \{y\}$ throughout Section 17.6.4.

\item p.~875, (17.47): The probability $P(s_j=0 \,| \ldots)$ is either 0 or 1, depending on whether the code bits $\{v_t\}$ together fulfill the check-sum $s_j$ or not. Therefore, (17.47) should be understood as
\eq{
\sigma_{j,l}^{x,(i)} = \sum_{\bs{b}} q_{j,t_1}^{b_1,(i)} \cdots q_{j,t_m}^{b_m,(i)}
}
where $\{t_1,\ldots,t_m\} = B(\bs{h}_j)\setminus \{l\}$ and the summation is over all even-weight (for $x=0$) or odd-weight (for $x=1$) binary $m$-tuples $\bs{b} = (b_1,\ldots,b_m)$.

\item p.~876, line 11: $\alpha_l^i$ should be $\alpha_l^{(i)}$.

\item p.~947, Problem~17.22: $m=6$ should be $k=6$.

\item p.~948, ref.~10: 432 should be 431.

\item p.~948, ref.~17: ``More'' should be ``New Results''.

\item p.~948, ref.~19: The first two authors, Kou and Lin, should be swapped.

\item p.~1102, ref.~17: The author order should be Lin, Rajpal, and Rhee, the book title should be \emph{Information Theory and Applications,} and the editors should be T.~A. Gulliver and N.~P. Secord.

\item p.~1102, ref.~18: The author order should be Lin, Rajpal, and Rhee, the editors should be R.~E. Blahut, D.~J. Costello, Jr., U.~Maurer, and T. Mittelholzer, and the city should be Boston.

\item p.~1250: Hocquengham should be Hocquenghem twice.

\end{itemize}

\end{document}